\documentclass[11pt,twoside]{article}
\usepackage{asp2010}
\usepackage{wasysym}

\resetcounters

\begin{document}

\title{Top-heavy IMFs in ultra-compact dwarf galaxies?}
\author{J. Dabringhausen$^1$, and P. Kroupa$^1$
\affil{$^1$Argelander-Institut f\"ur Astrophysik, Auf dem H\"ugel 71, 53121 Bonn, Germany}}

\begin{abstract}
Ultra compact dwarf galaxies (UCDs) are dense stellar systems at the border between massive star-clusters and small galaxies. Their on average high optical mass-to-light (M/L) ratio cannot be explained by stellar populations with the canonical stellar initial mass function (IMF), while it is doubtful that non-baryonic dark matter can accumulate enough on the scales of UCDs for influencing their dynamics significantly. UCDs in the Virgo galaxy cluster apparently also have an over-abundance of neutron stars, strongly suggesting a \emph{top-heavy} IMF, which would explain both findings. This is because a top-heavy IMF can provide the unseen mass through an abundance of stellar remnants. The suggested variation of the IMF can be understood if UCDs represent a case of rapid star-formation in an extremely dense environment. While top-heavy IMFs imply a much heavier mass-loss shortly after the formation of a stellar system, this process does not necessarily dissolve the UCDs. Their formation with a top-heavy IMF would therefore not contradict their existence.
\end{abstract}

\section{Introduction}
\label{intro}

Ultra-compact dwarf galaxies (UCDs) are stellar systems with $V$-band luminosities between $10^6$ and some $10^7$ Solar luminosities, but half-light radii of only about 50 pc or less. Most confirmed UCDs are found in nearby galaxies with rich globular cluster systems, such as NGC 5128 (Centaurus A), or nearby galaxy clusters, such as the Fornax cluster and the Virgo cluster. Even with the most powerful telescopes available at present, UCDs are barely resolved at those distances. Their true nature has therefore only been revealed quite recently by estimating their distance from their redshift \citep{1999A&AS..134...75H,2000PASA...17..227D}. Before that, UCDs were either interpreted as background galaxies or foreground stars, because of their appearance as bright point sources in most observations. Colours and in some cases also line indices suggest intermediate- to high ages for UCDs.

Due to the distance and the compactness of UCDs, it is difficult to get information on their internal stucture. The observed density profiles of UCDs carry a strong imprint of the point-spread function and the spectrum of a UCD is obtained in observations that cover a good fraction of the whole UCD. Estimates of their central velocity dispersions, their global velocity dispersions, their core radii or their half-light radii therefore require involved modelling of their density profiles, as done by, e.g. \citet{2007A&A...463..119H}. 

\begin{figure}
\plottwo{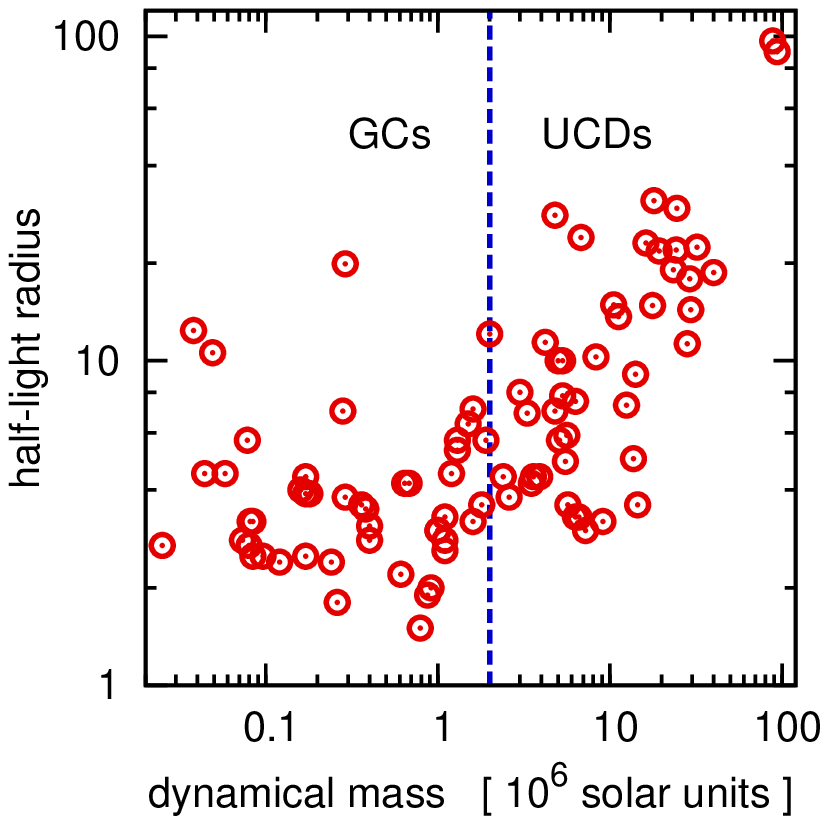}{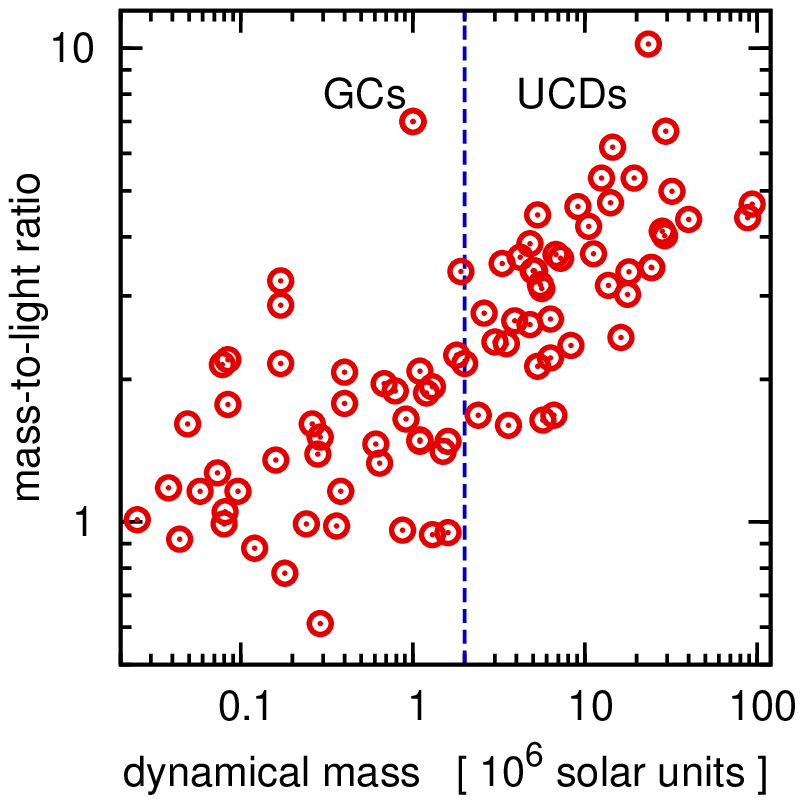}
\caption{Half-light radii $r_{\rm{h}}$ (left panel) and dynamical $M/L$ ratios (right panel) of GCs and UCDs, using the compilation of such objects by \citet{2008A&A...487..921M}. The dashed vertical lines at a dynamical mass, $M$, of $2 \times 10^6 \, M_{\odot}$ mark a definition that seperates GCs from UCDs, because objects with masses higher than $2 \times 10^6 \, M_{\odot}$ show a clear tendency to higher $r_{\rm{h}}$ and $M/L$ ratios, even though GCs are indistinguishable from UCDs at $M \approx 2 \times 10^6 \, M_{\odot}$. Note that also the two-body relaxation times become longer than a Hubble time $M \approx 2 \times 10^6 \, M_{\odot}$, i.e. GCs are relaxed stellar systems and UCDs are not.}
\label{fig:UCD_properties}
\end{figure}

A comprehensive list of globular clusters (GCs) and UCDs for which such modelling has been done, leading also to estimates for their dynamical masses, $M$, is found in \citet{2008A&A...487..921M}. The half-light radii ($r_{\rm h}$) and dynamical mass-to-light ($M/L$) ratios of the objects in that list are shown in Fig.~\ref{fig:UCD_properties}. According to this figure, UCDs are indistinguishable from GCs at a a luminosity of $\approx 10^6 \, L_{\odot}$, corresponding to a dynamical mass of $2 \times 10^6 \, M_{\odot}$ (see also \citealt{2008A&A...487..921M}). This has led some authors to consider UCDs as the most massive GCs (e.g. \citealt{2002A&A...383..823M}) or to re-interpret $\omega$ Cen, the most massive GC of the Milky Way, as a low-mass UCD. However,  the average $r_{\rm h}$) and average dynamical ($M/L$) ratios of objects with $M \apprge 2 \times 10^6 \, M_{\odot}$ increase systematically with $M$, whereas this is not the case for objects with $M < 2 \times 10^6 \, M_{\odot}$. This motivates to \emph{define} the objects with $M \ge 2 \times 10^6 \, M_{\odot}$ as UCDs, in order to distinguish them from 'classical' GCs.

The origin of UCDs is still a debated issue. However, numerical experiments predict that the young and massive star-cluster complexes observed in interacting galaxies evolve into objects very much like a UCD. The time required for this is only a few 100 Myr \citep{1998MNRAS.300..200K,2002MNRAS.330..642F}. Supporting evidence for this notion is provided by the discovery of W3, an object which was probably created by the formation of the merger-remnant galaxy NGC~7252 and which is a UCD in terms of its mass and its half-light radius \citep{2004A&A...416..467M,2005MNRAS.359..223F}.

\section{The M/L ratios of UCDs}
\label{MLratio}

Even if GCs and UCDs have the same origin, their differences in $M$ and $r_{\rm h}$ imply that they have evolved differently, which has implications on the difference of the average $M/L$ ratio between GCs and UCDs. The key to understanding this is that each stellar system is subject to two kinds of evolution: Stellar evolution and dynamical evolution. Stellar evolution implies that the brightest stars evolve first and thereby become dark remnants, leading to an increase of the $M/L$ ratio of a stellar system, unless they are ejected. Dynamical evolution, on the other hand, tends to lower the $M/L$ ratio of a stellar system, as it leads to a preferential loss of low-mass stars, which have a high $M/L$ ratio \citep{2003MNRAS.340..227B}. It is evident that the time-scale on which stellar evolution changes the properties of a stellar system is independent of $M$ and $r_{\rm h}$ of the stellar system, but this is not the case for dynamical evolution. A measure for whether a stellar system is affected by dynamical evolution is its median two-body relaxation time, $t_{rh}$, which can be estimated using equation~(2-63) from \citet{1987degc.book.....S}. If $t_{rh}$ is longer than the age of the system, it can be considered unaffected by \emph{dynamical} evolution.

Using their present-day parameters, it turns out that  GCs usually have $t_{\rm rh}$ very much below a Hubble time, whereas UCDs have $t_{\rm rh}$ of the order of a Hubble time or longer \citep{2008MNRAS.386..864D}. The $M/L$ ratios of GCs are thereby determined by an interplay of their stellar evolution as well as their dynamical evolution (see \citealt{2009A&A...500..785K}). UCDs are, in comparison, rather simple systems, as the evolution of their $M/L$ ratio is essentially determined by stellar evolution. Therefore, their $M/L$ ratio has only increased after their star formation came to an end. Thus, as the stellar populations of GCs and UCDs generally are of the same age (with the exception of W3),  the average $M/L$ ratio of UCDs being higher than that of GCs appears natural.

As UCDs are thus nearly unaffected by dynamical evolution, their present-day stellar population should be determined by their IMF and their star formation history. A good \emph{initial} assumption for the IMF of UCDs is the \emph{canonical} IMF, since this IMF is found to be consistent with all currently resolved stellar populations \citep{2001MNRAS.322..231K,2008LNP...760..181K}. However, the observed dynamical $M/L$ ratios of a clear majority of UCDs exceed the predictions for a single-age, single-metallicity stellar population (SSP) with the canonical IMF (see e.g \citealt{2008MNRAS.386..864D} or \citealt{2008A&A...487..921M}). This is even the case if  the stellar populations of all UCDs are assumed to be 13 Gyr old SSPs, i.e. if all stars in UCDs were nearly as old as the Universe. Note that for a given IMF and metallicity, no stellar population can have a higher $M/L$ ratio than this oldest possible stellar population. Also note that UCDs may consist of multiple stellar populations, as is the case with $\omega$~Cen, and that the sub-populations in them may have different metallicities, as  well as different ages. However, the metallicity of a UCD is estimated from its \emph{integrated} light, leading to some value that should be representative for all stars in the UCD.

\section{Top-heavy IMFs in UCDs?}
\label{top-heavy}

\subsection{Evidence for a top-heavy IMF in UCDs from their $M/L$ ratios}
\label{top-heavy-ML}

\begin{figure}
\plottwo{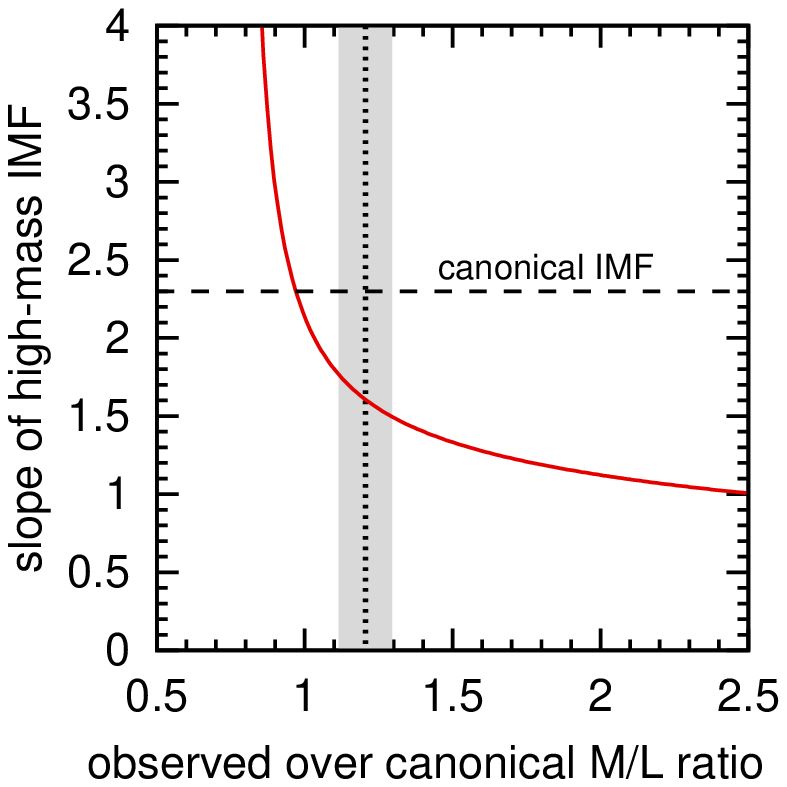}{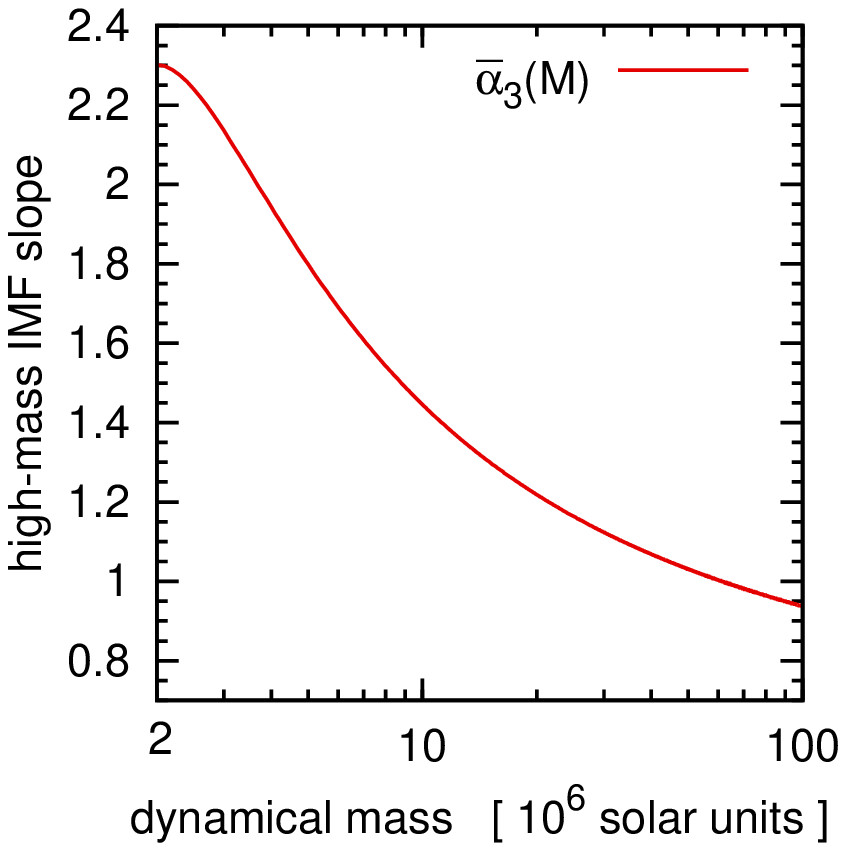}
\caption{The high-mass IMF slopes of UCDs (cf. Eq.~\ref{eq:IMF}), as suggested by their M/L ratio as a function of their observed over their predicted M/L ratio, assuming the canonical IMF (left panel), and as a function of their dynamical mass (right panel). For the curve in the left panel, it is assumed that 20 per cent of the NSs are retained, which seems to be a reasonable assumption (see Section~\ref{top-heavy-ML}). The average value for the M/L ratio of UCDs over the prediction for the canonical IMF (dotted vertical line) with its 1-$\sigma$ variance (shaded area) suggests that the IMF in UCDs is considerably top-heavy, even though the assumtion that they all are 13 Gyr old is very conservative.}
\label{fig:alpha_ML}
\end{figure}

The surprisingly high $M/L$-ratio of UCDs implies some unseen mass in them. This cannot be the non-baryonic dark matter that is often invoked to explain, e.g., the rotation curves of spiral galaxies or the $M/L$-ratio of galaxy clusters. This is because structures of cold dark matter have been shown to obey a universal density profile \citep{1997ApJ...490..493N,2001MNRAS.321..559B}. This density profile implies that only a very small amount of cold dark matter gathers within the quite compact UCDs and their $M/L$-ratios sould therefore be indistinguishable from the ones of pure stellar populations \citep{2009ApJ...691..946M}. The internal accelerations of UCDs are also above the transition between pure Newtonian dynamics and MOND, which is a modification of the classical theory of gravitation in the limit of weak gravitational fields \citep{1983ApJ...270..365M}. Thus, also MOND cannot explain the $M/L$-ratios of UCDs.

Excluding those possiblities leads to the assumption of an IMF that is \emph{not} universal, but varying, as an explanation for the high $M/L$ ratios of UCDs. This notion of a varying IMF does not neccesarily contradict the results by \citet{2001MNRAS.322..231K}, as these results were obtained considering resolved stellar populations in the Milky Way; i.e. from stellar populations in our immediate neighbourhood. UCDs, however, may have formed in much stronger starbursts than the ones that have occured in the Local Group. These extreme initial conditions would then be what distinguishes UCDs from the open clusters and GCs in the Milky Way.  It has indeed been shown that extremely high densities would lead to stellar collisions, which would alter the IMF \citep{2002MNRAS.336..659B}.

A varying IMF can be formulated as
\begin{equation}
\xi (m_*) =k k_i m_*^{-\alpha_i},
\label{eq:IMF}
\end{equation}
with 
\begin{eqnarray}
\nonumber \alpha_1 = 1.3,  & \qquad & 0.1  \le  \frac{m_*}{\rm{M}_{\odot}} < 0.5,\\
\nonumber \alpha_2 = 2.3,  & \qquad & 0.5  \le  \frac{m_*}{\rm{M}_{\odot}} < 1, \\
\nonumber \alpha_3 \, \in \, \mathbb{R}, & \qquad & 1 \le  \frac{m_*}{\rm{M}_{\odot}} \le 150,
\end{eqnarray}
where $m_*$ is the initial stellar mass, the factors $k_i$ ensure that the IMF is continuous where the power changes and $k$ is a normalisation constant \citep{2008LNP...760..181K}. $\xi(m_*)$ equals 0 if $m_*<0.1 \, \rm{M}_{\odot}$ or $m_*> 150 \, \rm{M}_{\odot}$, where $150 \, \rm{M}_{\odot}$ arguably is the upper mass-limit for stars (e.g. \citealt{2004MNRAS.348..187W,2005ApJ...620L..43O}). Eq.~\ref{eq:IMF} is the canonical IMF for $\alpha_3 = 2.3$. For $\alpha_3 < 2.3$, the IMF is top-heavy, implying more intermediate-mass stars and in particular more high-mass stars. In old stellar populations like the UCDs, these stars have evolved into white dwarfs and neutron stars (NSs) and thereby contribute to the mass of the total stellar population but hardly to its luminosity. We note that the IMF-variation proposed here is not the only one that can explain high M/L ratios. Alternative scenarios have been discussed by \citet{2008ApJ...677..276M} and \citet{2009ApJ...691..946M}. However, the advantage of the top-heavy IMF used here is that it can explain several properties of UCDs at the same time.

\begin{figure}
\plottwo{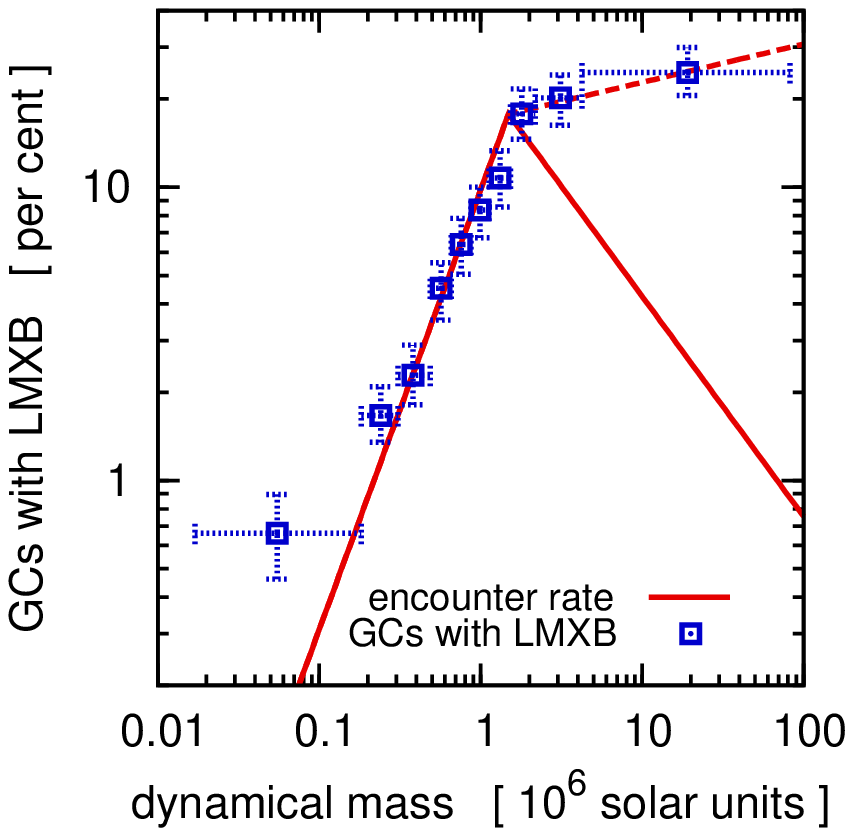}{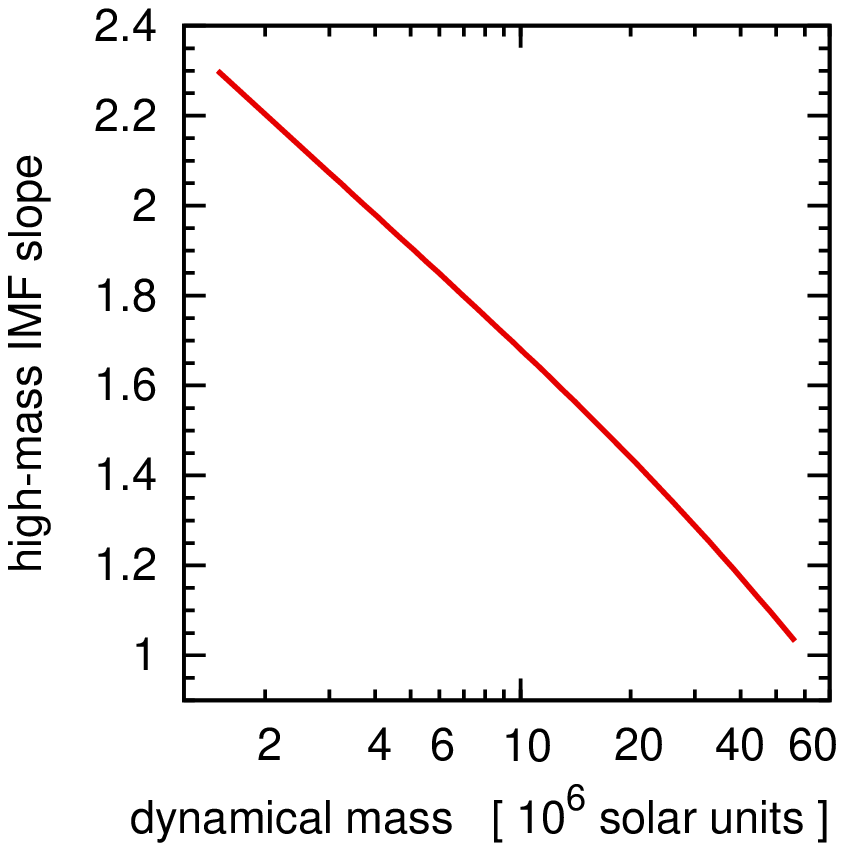}
\caption{Comparison of the encounter rate to the fraction of GCs and UCDs in the Virgo galaxy cluster containing an LMXB (left panel) and the high-mass IMF slope that can explain the abundance of LMXBs in UCDs (right panel). The x-axes in both panels have been re-scaled, so that they show the average dynamical mass corresponding to a certain luminosity instead of the luminosity itself. This allows easy comparisons between this figure and Figs.~\ref{fig:UCD_properties} and~\ref{fig:alpha_ML}. The solid line in the left panel shows the encounter rate under the assumption that the IMF in GCs and UCDs is canonical. By assuming that the IMF in GCs is canonical, but given by Eq. \ref{eq:IMF} in UCDs, the encounter rate in UCDs can by changed into the one indicated by the dashed line. This requires a certain dependency of the high-mass IMF slope on the mass of the UCDs. This function is shown in the right panel. It is remarkably similar to the one shown in the right panel of Fig.~\ref{fig:alpha_ML}, even though it was derived entirely differently.}
\label{fig:alpha_LMXB}
\end{figure}

Given Eq.~\ref{eq:IMF}, it is possible to translate some observed $M/L$ ratio divided by the $M/L$-ratio predicted by a SSP-model into some value of $\alpha_3$ that corresponds to this ratio by providing the appropriate amount of dark matter in form of WDs and NSs (see left panel of Fig.~\ref{fig:alpha_ML}). Note that the fraction of NSs that actually remain bound to the UCD is an important issue when calculating $\alpha_3$. Given the large peculiar velocities of many pulsars \citep{1994Natur.369..127L}, which are thought to be NSs, a lot of NSs may escape from the UCDs. On the other hand, it is observationally confirmed that GCs and especially UCDs do have a population NSs (visible as low-mass X-ray binaries, cf. Section \ref{top-heavy-LMXB}).

The right panel of Fig.~\ref{fig:UCD_properties} shows that there is a relation between the $M/L$ of UCDs and their dynamical mass. Combining this information with the $\alpha_3$ implied by a certain $M/L$ ratio (left panel of Fig.~\ref{fig:alpha_ML}) leads to a relation between the average $\alpha_3$ of UCDs as a function of their mass (see \citealt{2009MNRAS.394.1529D} for details). This relation, as shown in fig.~5 in \citet{2009MNRAS.394.1529D}, is plotted in the right panel of Fig.~\ref{fig:alpha_ML}.

\subsection{Evidence for a top-heavy IMF in UCDs from their low-mass X-ray binaries}
\label{top-heavy-LMXB}

If UCDs have indeed formed with the IMF formulated in Eq.~\ref{eq:IMF} and the IMF is top-heavy (i.e. $\alpha_3<2.3$), they should have a large population of NSs. Thus, compared to the case with the canonical IMF, UCDs would also have a higher number of close binary systems of a low-mass star and a NS. In such systems, the NS can accrete matter from its companion star and thereby become a bright X-ray source; a so-called low-mass X-ray binary (LMXB). The NSs in LMXBs thereby become observable, even though they remain invisible in the optical.

Observations show that the densest stellar systems also have the highest likelihood to harbour a bright LMXB. This suggests that they are formed by three-body encounters; especially encounters between a star and a binary. This is because these encounters tighten binary systems if the binary entering the interaction was already tight \citep{1975MNRAS.173..729H}. It thus appears that stellar dynamics is the most important and perhaps even the only important factor for the formation of LMXBs \citep{2003ASPC..296..245V,2007ApJ...660.1246S,2010MNRAS.tmp.1159P}.

The number of encounters relevant for the creation of LMXBs, i.e. encounters that involve stars and NSs, can be written as
\begin{equation}
\Gamma \propto \frac{n_{\rm ns}n_{\rm s} r_{\rm c}^3}{\sigma},
\label{eq:Gamma}
\end{equation}
where  $n_{\rm ns}$ is the number density of NSs, $n_{\rm s}$ is the number density of other stars, $r_{\rm c}$ is the core radius and $\sigma$ is the velocity dispersion  \citep{2003ASPC..296..245V}. Rather than measuring the volume of the stellar system by $r_{\rm c}^3$, we use $r_{\rm h}^3$. This is a quantity that is available for all GCs and UCDs plotted in Fig.~\ref{fig:UCD_properties} and arguably only changes the proportionality factor in Eq.~\ref{eq:Gamma}.

Interesting for the purpose here is to find the average encounter rate of GCs and UCDs as a function of their luminosity, $\overline{\Gamma}(L)$. This quantity can be compared to the fraction of GCs and UCDs in certain luminosity-bins that actually have an LMXB, $P$, which is an estimator for the probability to find an LMXB in a system in the according luminosity range (and a corresponding mass range). Observational data on which fraction of GCs and UCDs in the Virgo-galaxy cluster have a bright LMXB can be found in \citet{2007ApJ...660.1246S}. If the stellar mass function in GCs is the same as in UCDs, the shape of $\overline{\Gamma}$ only depends on how the average $r_{\rm h}$ of GCs and UCDs changes with their luminosity, which is a function that can be derived using the data plotted in Fig.~\ref{fig:UCD_properties}. A non-changing stellar mass function for all GCs does indeed suggest that the observed rise of $P$ matches the increase of $\overline{\Gamma}$ expected for them. However, $P$ still rises for UCDs, whereas the calculated $\overline{\Gamma}$ drops steeply if  they all had the same stellar mass function (cf. left panel of Fig.~{\ref{fig:alpha_LMXB}).

This leads to the alternative assumption that the mass function of UCDs \emph{does} change with their luminosity. An IMF that becomes increasingly top-heavy with the luminosity of UCDs implies that $n_{\rm s}$ in Eq.~\ref{eq:Gamma} changes with luminosity. Consequently, also $\sigma$ increases, because more NSs imply a higher total mass.  Assuming the population of NSs and other stars in GCs is given according to the canonical IMF and that the IMF in UCDs is given by Eq.~\ref{eq:IMF}, $\Gamma$ in UCDs can be calculated such that it is not only proportional to $P$ in GCs, but also in UCDs. This implies that that the high-mass slope of the IMF changes with their luminosity (mass), as shown in the right panel of Fig.~\ref{fig:alpha_LMXB}. Comparing the right panel of Fig.~\ref{fig:alpha_LMXB} with the right panel of Fig.~\ref{fig:alpha_ML} shows a qualitative agreement of how the high-mass IMF slope of UCDs would have to change in order to explain their $M/L$ ratios and how it would have to change in order to explain the abundance of LMXBs in them (Dabringhausen \& Kroupa, in preparation).

\section{Implications for the inital conditions in UCDs}
\label{initial}

The dynamical evolution of young stellar systems is strongly influenced by rapid mass-loss, especially if their IMF is top-heavy. This is because of the strong radiation of massive stars, which can quickly remove all the interstellar medium that is not used up in star formation.  Massive stars evolving to SN can in principle replenish the interstellar medium, but provide at the same time so much energy that also this gas is more likely to leave the stellar system, even if it is as massive a UCD. Stellar evolution can thus lead to further mass-loss that becomes more rapid and more severe with the the top-heaviness of the IMF. This mass-loss would lead to a strong expansion or even to a dissolution of the stellar system. The extreme IMFs suggested for UCDs in Sections~\ref{top-heavy-ML} and~\ref{top-heavy-LMXB} therefore imply that UCDs must have been extremely compact if they are to survive until today with the observed densities \citep{2010MNRAS.403.1054D}. Assuming that their stellar populations formed on a time-scale of a few Myr (as implied by the formation scenario suggested in Section \ref{intro}), the more massive UCDs may have had initial masses of some $10^8 \, M_{\odot}$ while having initial half-light radii of only a few parsec. A top-heavy IMF would then imply a population of the order of $10^6$ O-stars with a total luminosity of the order of $10^{11} \, L_{\odot}$. If this is the case, UCDs should be bright point-like sources, identifiable by a spectrum  that is consistent with a stellar population (Pflamm-Altenburg, Dabringhausen \& Kroupa, in preparation). 

The high initial density implied by the initial mass and initial radius could explain why the IMF was top-heavy in the first place. At such densities, collisions between proto-stars become common, which may cause a deviation from the canonical IMF \citep{2002MNRAS.336..659B}.

\section{Conclusion}
\label{conclusion}

The observed dynamical $M/L$ ratios of UCDs and the number of LMXBs in them can be used to constrain their IMF. Their high $M/L$ ratios are a strong argument for a non-canonical IMF in UCDs, but the $M/L$ ratios alone leave a number of possibilities on how the IMF could vary with the mass of the UCDs. However, UCDs apparently also have a richer population of NSs than GCs. This cannot be explained by simply assuming that more NSs remain bound to UCDs, while GCs and UCDs both formed with the canonical IMF, because in this case the $M/L$ ratios of UCDs should not tend to exceed the value predicted for a canonical IMF. Thus the number of LMXBs in UCDs and the $M/L$ ratios of UCDs in combination imply that the IMF of UCDs was top-heavy, especially for very high-mass stars. This implies that UCDs would have formed compact and with very high densities. Note that the number of LMXBs also give a constraint on the low-mass IMF in UCDs, because low-mass stars have to be rather frequent as well. Otherwise, LMXBs in UCDs would be rare for a lack of low-mass companions for the NSs (cf. Eq~\ref{eq:Gamma}).

\bibliography{Dabringhausen_J}

\end{document}